\documentclass[12pt]{article}
\usepackage{graphicx}
\date{}
\def\be{\begin{equation}}
\def\ee{\end{equation}}
\def\bea{\begin{eqnarray}}
\def\eea{\end{eqnarray}}
\def\s{\sigma}

\def\lm{\lambda}
\def\de{\delta}
\def\om{\omega}
\def\pr{\prime}
\def\f{\varphi}
\def\th{\theta}

\def\tom{\tilde\omega}

\title{STABILITY AND SYMMETRY BREAKING \\
FOR CLOSED STRING WITH MASSIVE POINT\thanks{Is supported
by Russian foundation of basic research, grant 05-02-16722}\\
}

\author{A.\,E. Milovidov, G.\,S. Sharov\\
{\small Tver state university}\\
{\small Tver, 170002, Sadovyj per. 35, Mathem. dep-t}}
\begin{document}
\maketitle
\begin{abstract}
The closed relativistic string carrying a point-like mass in the space
with nontrivial geometry is considered. For rotational states of this system
(resulting in non-trivial Regge trajectories) the stability problem
is solved. It was shown that rotations of the folded string with the
massive point placed at the rotational center are stable (with respect
to small disturbances) if the mass exceeds some critical value: $m>m_{cr}$.
But these rotational states are unstable in the opposite case $m<m_{cr}$.
We can treat this effect as the spontaneous symmetry breaking
for the string state.
Other classes of rotational motions of this system have appeared to be stable.
These results were obtained both in numerical
experiments and the analytical investigation of small disturbances
for the rotational states.
\end{abstract}

\centerline {\bf 1. Introduction}
\medskip

We consider the closed relativistic string carrying one massive
point. This system moves in the space ${\cal M}=R^{1,3}\times T^{D-4}$, where
$R^{1,3}$ is $3+1$\,-\,dimensional Minkowski space and the compact manifold
$T^{D-4}$ is $D-4$\,-\,dimensional torus resulting from the compactification
procedure \cite{GSW}. The structure of homotopic classes for the closed
string in the manifold ${\cal M}$ is non-trivial (except for the simplest variant
${\cal M}=R^{1,3}$, $D=4$,
but it is also included as the particular case into our consideration).

The action of this system is \cite{MilSh}
\be
S=-\gamma\int\limits_{\Omega}\sqrt{-g}\;d\tau\,d\s
-m\int\sqrt{\dot x_1^2(\tau)}\;d\tau. \label{S}
\ee

Here $\gamma$ is the string tension, $m$ is the point-like mass,
$g$ is the determinant of the induced metric
$g_{ab}=G_{\mu\nu}(X)\,\partial_a X^\mu\partial_b X^\nu$
on the string world surface $X^\mu(\tau,\s)$, embedded in the manifold
${\cal M}=R^{1,3}\times T^{D-4}$. We denote $x^0,x^1,x^2,x^3$ the coordinates
in Minkowski space $R^{1,3}$ and the torus $T^{D-4}$ has
cyclic coordinates $x^k$ ($k=4,5\dots$) with periods $\ell_k$,
that is, points with coordinates $x^k$ and $x^k+N_k\ell_k$,
$N_k\in Z$ are identified.
We suppose $T^{D-4}$ to be flat manifold so the metric of ${\cal M}$
is equal to that of Minkowski space $R^{1,D-1}$:
$G_{\mu\nu}(X)=\eta_{\mu\nu}$.

In the action (\ref{S}) the speed of light $c=1$,
$\Omega=\{\tau,\s:\,\tau_1<\tau<\tau_2,\,\s_1(\tau)<\s<\s_2(\tau)\}$;
the equations $x_i^\mu(\tau)=X^\mu(\tau,\s_i(\tau))$, $i=1,2$
describe the same trajectory of the massive point
\be
x_1^\mu(\tau)=X^\mu(\tau,\s_1(\tau))=X^\mu(\tau^*,\s_2(\tau^*))+
\sum_k N_k\ell_k\de^\mu_k\
\label{close}
\ee
on the tube-like world surface \cite{Tr,PRTr}.
Here $\de^\mu_k=\left\{\begin{array}{cl} 1, & \mu=k,\\ 0, & \mu\ne k.
\end{array}\right.$
This line can be parameterized  with two different parameters $\tau$ and
$\tau^*.$ They are connected by the relation $\tau^*=\tau^*(\tau).$
This relation
should be added to the closure condition (\ref{close}) of the world surface.

Equations of motion for this system
result from the action (\ref{S}) via its variation.
They may be reduced to the simplest form \cite{MilSh} without loss of
generality under the orthonormality conditions on the world surface
\be
(\partial_\tau X\pm\partial_\s X)^2=0,
\label{ort}\ee
and the conditions
\be
\s_1(\tau)=0,\qquad \s_2(\tau)=2\pi.
\label{ends}\ee
Under these conditions the string equations of motion
take the mentioned simplest form
\be
\frac{\partial^2X^\mu}{\partial\tau^2}- \frac {\partial^2X ^\mu}
{\partial\s^2}=0,\label{eq}
\ee
\be
m\frac d {d\tau}\frac{\dot X^\mu (\tau, 0)}{\sqrt{\dot X^2(\tau, 0)}}+
\gamma\big[X^{'\!\mu}(\tau^*,2\pi)-X^{'\!\mu}(\tau,0)\big]=0.
\label{mg}
\ee
Here $\dot X^\mu\equiv\partial_\tau X^\mu $,$X^{'\!\mu}\equiv
\partial_\s X^\mu$; scalar square is
$a^2=\eta_{\mu\nu}a^\mu a^\nu $. Equation (\ref{mg}) may be treated
as the boundary condition for Eq.~(\ref{eq}).

Thus, dynamics of the closed relativistic string with the massive point
in the manifold ${\cal M}$ is described by the system (\ref{close})\,--\,(\ref{mg}).

The solutions of system (\ref {close})\,--\,(\ref{mg}) was obtained in
Ref.~\cite{MilSh} under the restrictions
\be \frac\gamma {m}
\sqrt{\dot X^2 (\tau, 0)}=Q={}\mbox {const},\qquad\tau^*
=\tau+\tau_0,\qquad\tau_0={}\mbox{const}\label{Ctau0}
\ee
in the form of the following Fourier series:
\be X^\mu=\sum\limits_{n=-\infty}^\infty \chi_n^\mu(\s) \exp(-i\om_n\tau).
\label{Four}
\ee
Frequencies $\om_n$ (in pairs with values $\tau_0$) are solutions of
system of the equations
\be
\cos2\pi\om_n-\cos\om_n\tau_0=\frac{\om_n}{2Q}
\sin2\pi\om_n,\label{con}
\ee\vspace{-3mm}
\be 1+\frac{\tau_0^2}{4\pi^2}-2\frac {\tau_0}{2\pi}
\frac{1-\cos\om_n\tau_0\cdot\cos2\pi\om_n}
{\sin\om_n\tau_0\cdot\sin2\pi\om_n} = \left (1-\frac{\tau_0}
{2\pi}\frac{\sin2\pi\om_n}{\sin\om_n\tau_0}\right)\frac
{\gamma^2}{m^2Q^2}\sum_{k>3} b_k^2.
\label{omtaub}\ee
Here $\displaystyle b_k =\frac {\ell_k N_k}{2\pi}$ are the factors
connected with winding numbers $N_k$ for the cyclic coordinates $x^k$, $k\ge4$.

We shall consider below the so called one-frequency states. They are
solutions of the type (\ref{Four}) containing only one
nonzero frequency $\om_n$ and describing rotational motions of the system.
These states are the most interesting ones, because they generate
non-trivial spectrum of Regge trajectories \cite{MilSh} and may be
applied in the hadron spectroscopy.

One-frequency states of the string system (\ref{S}) may be divided
into three classes. In the case $\tau_0\ne0$ they have the form \cite{MilSh}
\be
\begin {array} {c} \displaystyle
X^\mu=e_0^\mu a_0\Big(\tau-\frac{\tau_0}{2\pi}\s\Big)
+\sum\limits_{k> 3}e^\mu_k b_k\s +\\
+A_n\Big\{\big[S\cos\om_n\s+(C_0-C)\sin\om_n\s\big]\cdot e
^\mu (\om_n\tau)-S_0\sin\om_n\s\cdot\acute e ^\mu (\om_n\tau)
\Big\}.\rule [3mm]{0mm}{1mm}
\end {array}
\label{hyp}
\ee
Here $e_0,\,e_1, \,e_2,\,\dots\, e_{D-1}$ are vectors of some fixed
orthonormal basis in the manifold ${\cal M}$
($e_0$, $e_1$, $e_2$, $e_3$ is the basis in $R^{1,3}$) and
\be
e^\mu(\om_n\tau)=e^\mu_1\cos\om_n\tau+e^\mu_2\sin\om_n\tau, \qquad
\acute e^\mu(\om_n\tau)=\om_n^{-1}\frac d{d\tau} e^\mu(\om_n\tau)
\label{eacute}
\ee
are unit orthogonal rotating vectors.
The following notations are used
\be
 C =\cos2\pi\om_n, \qquad S=\sin2\pi\om_n,
\qquad C_0 =\cos\om_n\tau_0, \qquad S_0=\sin\om_n\tau_0,
\label{cscs}
\ee
where the values $\om_n$ and $\tau_0$ are taken from Eqs.~(\ref{con}),
(\ref{omtaub}).

Parameters $a_0$, $A_n$ are related by the conditions
\be
a_0 =\frac{mQ}\gamma(1-v^2)^{-1/2}, \qquad
A_n =\frac{a_0v}{\om_nS},\qquad v^2 =\frac{\tau_0S}{2\pi S_0}.
\label{a0A}
\ee
Here $v$ is the speed of the massive point moving along a circle;
$0\le v<1$.

Solutions (\ref{hyp}) describe uniform rotations of
the closed string. In the case $b_k=0$ the string has the form of a closed
hypocycloid joined at non-zero angle in the massive point. This
hypocycloid uniformly rotates in the $e_1,e_2$ plane.
The similar solutions for the string baryon model ``triangle''
were obtained and studied in Ref.~\cite{Tr}

In the case $\tau_0=0$ rotational states of the system are divided into two
classes, because the equation (\ref{con}) (defining their frequencies $\om_n$)
has the form
\be
\sin\pi\om_n\cdot\Big (\sin\pi\om_n +\frac
{\om_n}{2Q}\cos\pi\om_n\Big) =0.
\label{q}
\ee
The roots $\om_n=-2Q\tan\pi\om_n$ of Eq.~(\ref{q})
determine rotational motions in the form

\be X^\mu=e_0^\mu a_0\tau
+\sum_{k>3}e^\mu_k b_k\s+A_n\cos\big[\om_n (\s-\pi)\big]\cdot e
^\mu (\om_n\tau).\label{rot}
\ee
Solutions (\ref {rot}) describe
uniform rotations of the sinusoidal string. The massive point with
mass $m$ (at $\s=0$) rotates with the string. The trajectory of this point
is a circle.

The roots of the equation (\ref{q}) $\om_n=n$ result in the following solutions:
\be
X^\mu=e_0^\mu a_0\tau+\sum_{k> 3} e^\mu_k b_k\s+
A_n\sin n\s\cdot e^\mu(n\tau),\label{sl}
\ee

Here $a_0=\sqrt{n^2A_n^2+\sum b_k^2}$. These solutions
describe rotations of the string similar to motions (\ref{rot})
but the massive point for states (\ref{sl}) is placed at the rotational center.

In the case $b_k=0$ expressions (\ref{rot}) and (\ref{sl}) describe
uniform rotations of the string folded several times.
It rotates in Minkowski space $R^{1,3}$ and the cyclic coordinates of
$T^{D-4}$ do not ``work''.
Such string motions have been classified in Ref.~\cite{PeSh}.

For rotational motions (\ref{hyp}), (\ref{rot}) and (\ref{sl}) in the case
$b_k=0$ the string has points moving at the speed of light.
There are singularities of the metrics $\dot X^2=X^{'2}=0$ at the world
lines of these points on the world surface. At the same time,
for $b_k\ne0$ the mentioned solutions have no such singularities.

Possible applications of the solutions (\ref {hyp}), (\ref {rot}) and
(\ref {sl}) in hadron spectroscopy essentially depend on stability
or instability of these states with respect to small disturbances.
In the following section the numerical experiments in this direction are
described. They showed rather unexpected results and in Sect.~3 the
stability problem for the mentioned rotational states is studied analytically.

\bigskip \bigskip

\centerline {\bf 2. Numerical description and visualization of arbitrary motions
of the system}
\medskip

To solve the stability problem for rotational motions (\ref{hyp}), (\ref{rot})
and (\ref{sl}) we study evolution of their small disturbances numerically.

Any motion of the system with the action (\ref{S}) can be determined
unambiguously, if the following initial conditions of the system are given:
an initial position of the system in Minkowski space
and initial velocities of string points \cite{stabPRD}.
The solution of this initial-boundary value problem for various string models
meson and baryons was suggested earlier in Refs.~\cite{stabPRD,Y02}.

We apply this approach to the closed string with the action (\ref{S})
using the general solution of the equation of motion (\ref{eq})
\be
X^\mu (\tau,\s)=\frac{1}{2}[\Psi^\mu_{+}(\tau +\s)+\Psi
^\mu_{-}(\tau-\s)].\label{gensol}
\ee
Here
\be{\Psi'_+\!\!}^2={\Psi'_-\!\!}^2=0
\label{Psi2}
\ee
by virtue of the orhthonormality conditions (\ref{ort}).

It is convenient to use the following notation for the unit vector of
velocity of the massive point
$$U^\mu=\frac{\dot X^\mu(\tau,0)}{\sqrt{\dot X^2(\tau,0)}}=
\frac{\dot X^\mu (\tau^*,2\pi)}{\sqrt{\dot X^2(\tau^*,2\pi)}}.
$$

It takes the form
\be U^\mu=\frac{\Psi'^\mu_{+}\tau)+
\Psi'^\mu_{-}(\tau)}{\sqrt{2\langle\Psi'_{+}(\tau),
\Psi'_{-}(\tau)\rangle}}=\frac{\Psi'^\mu_{+}(\tau^*
+2\pi)+\Psi'^\mu_{-}(\tau ^*-2\pi)}
{\sqrt{2\langle\Psi'_{+}(\tau^*+2\pi),\Psi'_{-}(\tau^*-2\pi)\rangle}}.
\label{U}\ee
after substitution of the expression (\ref{gensol}).
Here and velow the square product is denoted
$\langle a,b\rangle=\eta_{\mu\nu}a^\mu b^\nu $.

We reduce the equations of motion (\ref {close})\,-\,(\ref{mg})
to the system of differential equations with respect to
functions $\Psi^\mu_\pm$, $U^\mu$. For this purpose we
multiply the expression (\ref{U}) by vectors $\Psi'_{\pm}(\tau)$,
$\Psi'_{\pm}(\tau^*\pm2\pi)$ and take into account the conditions (\ref {Psi2}).

We obtain the following relation:
$$\langle U,\Psi'_{\pm}(\tau)\rangle=
\sqrt{\frac{\langle\Psi'_{+}(\tau),\Psi'_{-}(\tau)\rangle} 2},
\quad\langle U,\Psi'_{\pm}(\tau^*\pm 2\pi)
\rangle=\sqrt{\frac{\langle\Psi'_{+}(\tau^*+2\pi),
\Psi'_{-}(\tau^*-2\pi)\rangle }2}.$$

This equality allows us to rewrite expressions (\ref{U}) as
 \be
\Psi'^\mu_{+}(\tau)+\Psi'^\mu_{-}(\tau)=2\langle
U,\Psi'_{\pm}(\tau)\rangle\, U^\mu,\label{vir1}
\ee
\be
\Psi'^\mu_{+}(\tau^*+2\pi)+\Psi'^\mu_{-}(\tau^*-2\pi)
=2\langle U,\Psi'_{\pm}(\tau^*\pm 2\pi)\rangle\, U^\mu.\label {vir2}
\ee
Using these relations, we write down the boundary condition (\ref{mg})
as follows:
\be
\frac{dU^\mu}
{d\tau}=-\frac{\gamma}{m}\Big [U^\mu(\tau)\,U_\nu(\tau)-
\de^\mu_\nu\Big ]\Big [\Psi'^\nu_{+}(\tau)+\Psi'^\nu_{-}
(\tau^*-2\pi)\Big ].\label{mgg}
\ee

Differentiating the closure condition (\ref{close}) on $\tau$
$$
\dot X^\mu (\tau^*,2\pi)\cdot\frac{d\tau^*}{d\tau}=
\dot X^\mu (\tau,0)$$
and squaring this equality with using Eqs.~(\ref{gensol}), (\ref {Psi2})
we obtain the relation between $\tau^*$ and $\tau$:
\be\frac{d\tau^*}{d\tau}=\frac{\langle U,\Psi'_{\pm}(\tau)\rangle }
{\langle U,\Psi'_{\pm}(\tau\pm 2\pi)\rangle }.\label{zavc}
\ee
Note that the function $\tau^*(\tau)$ in Eq.~(\ref{close}) can not be
taken in an arbitrary form. It should be calculated from the dynamical equations
of this system, in particular, from Eq.~(\ref{zavc}).

Equations (\ref {vir1})\,--\,(\ref{zavc}) allow us to describe numerically
any motion of the closed relativistic string with one point-like mass.

An initial position of the string can be given as the parametric curve
$$x^\mu=\rho^\mu(\lm),\qquad \lm\in[\lm_1, \lm_2]
$$
and initial velocities of string points as the function
$v^\mu (\lm)$, $\lm\in [\lm_1, \lm_2]$. Here $\rho'(\lm)$ is the
space-like vector, $v^\mu$ is the time-like vector.

Consider in more detail the numerical procedure for solving the described
initial-boundary value problem. This procedure includes three stages.

At the first stage it is necessary to calculate the functions
$\Psi_\pm^{\pr\mu}$ in the segments of influence of the initial data.
We can do it without loss of a generality if we use the freedom in choice of two functions $\check\tau(\lm)$ and $\check\s(\lm)$ in parametrization
of the initial curve (initial position of the string) \cite{stabPRD,Y02}:
$$X^\mu\big (\check\tau(\lm), \check\s(\lm)\big)=\rho^\mu (\lm).
$$
For the considered system it is convenient to choose them as
$$
\check\tau'(\lm)=\langle v,\rho'\rangle\big/v^2,\qquad
\check\s'(\lm)=\Delta\big/v^2,
$$
Here $\Delta(\lm)=\sqrt{\langle v,\rho'\rangle^2-v^2\rho^{\pr2}}$.

Then we integrate numerically $\check\tau'(\lm)$ and $\check\s'(\lm)$ using the
Simpson's method and obtain the functions $\check \tau(\lm)$ and
$\check\s(\lm)$, under the assumption $\check\tau(\lm_1)=\check\s(\lm_1)=0$.

With  the help of them it possible to find the functions
$\Psi_\pm^{\pr\mu}$ via formulas \cite{stabPRD}
$$\Psi_\pm^{\pr\mu}\big(\check\tau(\lm)\pm\check\s(\lm)\big)=\frac
{(\Delta\pm P)\rho^{\pr\mu}\mp\rho^{\pr 2}v^\mu}
{\Delta\big [\check\tau'(\lm)\pm\check\s'(\lm) \big]}$$
on initial segments
\be
\Psi_+^{\pr\mu}(\xi_+),\quad\xi_+\in[0;\check\tau(\lm_2)+\check\s(\lm_2)];\qquad
\Psi_-^{\pr\mu}(\xi_-),\quad\xi_-\in[\check\tau(\lm_2)-\check\s(\lm_2);0].
\label{pseg}\ee
They are the mentioned segments of influence of the initial data.

For the further numerical calculation it is convenient to recalculate
the values of functions $\Psi_\pm^{\pr\mu}$ in equidistant points of their
arguments with step $h$ using linear approximation.

The second stage includes extension of the functions $\Psi_\pm^{\pr\mu}(\tau)$
beyond the initial segments (\ref{pseg}) up to given maximal value of their argument $\tau_{max}$. For this purpose we integrate the equation (\ref{mgg})
and simultaneously calculate current values of functions
$\Psi_{-}^{\pr\mu}(\tau)$,
$\Psi_{+}^{\pr\mu}(\tau^*+2\pi)$ and $\tau^*(\tau)$
using the equations (\ref{vir1}), (\ref {vir2}), (\ref {zavc}).

To increase accuracy of the difference scheme we substitute
$\big [U^\mu(\tau+h)-U^\mu(\tau)\big]/h$ into the left part of Eq.~(\ref{mgg}) and calculate the value of the right part at $\tau+h/2$. For this reason functions $\Psi_\pm^{\pr\mu}(\tau)$ are taken at the points $\tau+h/2$.
Values of $U^\mu$ at these points are replaced by
$\frac12\big [U^\mu (\tau)+U^\mu (\tau+h)\big]$.

Eq.~(\ref{mgg}) will be transformed into the following difference scheme:
$$
U^\mu (\tau+h)\bigg (1+\frac{\gamma h}{2m}\langle U,\Pi\rangle\bigg)
=U^\mu(\tau)+\frac{\gamma h}{m}\bigg (\Pi^\mu-\frac{U ^\mu (\tau)}{2}
\langle U,\Pi\rangle\bigg).
$$
Here $\Pi ^\mu =\Psi_{+}^{\pr\mu}(\tau+h/2)+
\Psi_{-}^{\pr\mu}\big(\tau^*(\tau+h/2)-2\pi\big)$.

Using the obtained values $U^\mu (\tau+h)$ we calculate
$\Psi_{-}^{\pr\mu}(\tau+h/2)$ from Eq.~(\ref{vir1}),
$\Psi_{+}^{\pr\mu}(\tau^*+2\pi+h/2)$ from Eq.~(\ref {vir2}),
and $\tau^*(\tau)$ from Eq.~(\ref{zavc}.

When the values of $\Psi_{pm}^{\pr\mu}$, $\tau^*(\tau)$
are calculated from $\tau=0$ up to $\tau=\tau_{max}$,
we integrate the vector functions $\Psi_\pm ^{\pr\mu}$ with the help of
the Simpson's method (approximating the integrated function by square
trinomials).
If values of the function are known in half-integer points this
method is reduced to the formula
$$
\int\limits_0^h f(x)\,dx=\frac h {24}\bigg [f\Big (-\frac h2\Big )+
22f\Big (\frac h2\Big )+f\Big (\frac {3h}2\Big )\bigg ].
$$

At the third stage of this procedure we calculate the function
$X^\mu(\tau,\s)$ from Eq.~(\ref{gensol}) (the world surface) and
build projections of lines  $t=X^0(\tau,\s)={}$const
onto the $xy$ plane  where the axes $x$ and $y$ are directed accordingly
along vectors $e_1$ and $e_2$.

Visualization of string motions was represented in various ways,
in particular, as animated cartoons. In this case the second and third stages
of the calculations should be made almost simultaneously
(with certain delay).

\begin{figure}[hb]
\centering
\includegraphics[width=\textwidth]{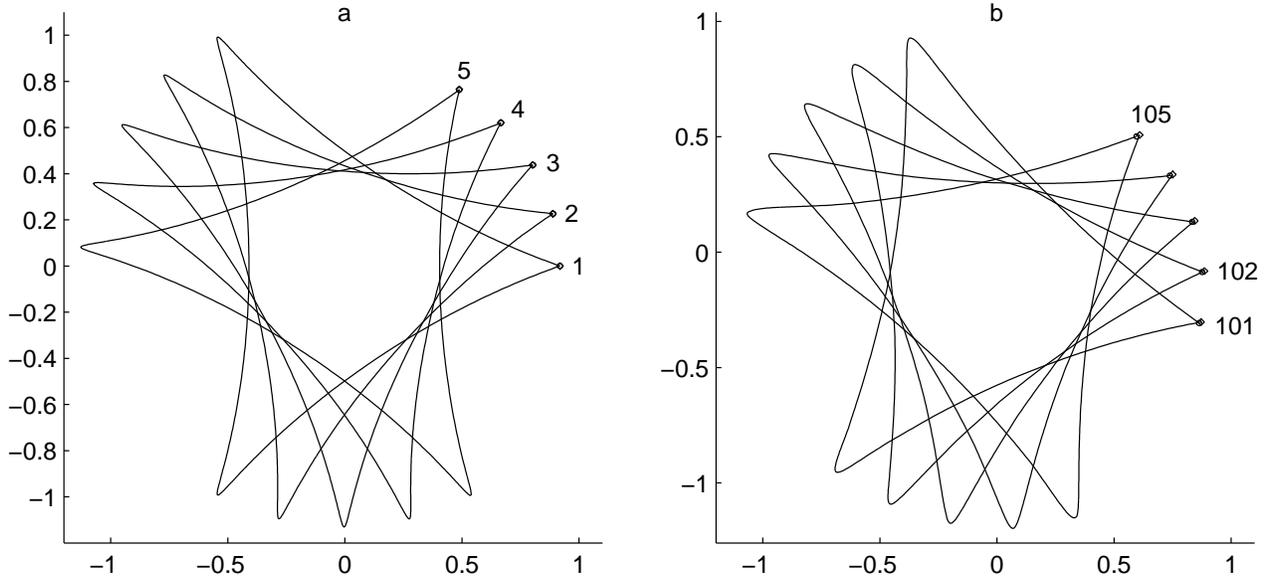} 
\caption{\small Quasirotational motion of the system (\ref
{hyp}).}
\end{figure}

Another way of visualization is shown below in Figs.~1 and 2. In
these figures projections of sequential positions of the system on
the plane $xy$ (projection of sections $t={}$const of the world
surface) are made at regular intervals of time $\Delta t=0{.}3$.
Numbers of these positions in ascending order of $t$ are
specified. Position of the massive point is designated by the
small circle.

Fig.~1 demonstrates the example of motion of the system close to the
exact solution (\ref {hyp}) with parameters
$m/\gamma=1$; $\sum b_k^2=0{.}25$; $a_0=1{.}55$; $\om=1{.}285$; $\tau_0=2{.}006$. These curves slightly differ from hypocycloids
because of $b_k\ne$.
The initial position (marked by number 1 in Fig.~1a) corresponds to the
section $t=0$ of the world surface (\ref{hyp}).
Initial velocity contains small disturbance in the form
$\de v^\mu=e_2^\mu\cdot\sin^2\s$.

Numerical experiments show, that amplitudes of small disturbances of states
(\ref{hyp}) and also (\ref{rot}) do not grow during enough large time
intervals $t$.
For example, in Fig.~1b the position 101 corresponds to $t=30$, the
curvilinear triangle after 3 rotations keeps its form in general.
We may conclude that the rotational states (\ref{hyp}) and also (\ref{rot})
are stable with respect to small disturbances on the level of
numerical simulation.

Numerical investigations of the rotational states (\ref{sl}) with the
massive point at the center of rotation showed interesting and unexpected
results. We obtained that stability of this state depends on
values of mass $m$ of the material point, or, more precisely, it depends on
values of parameter $m/\gamma$. If this value is large enough, the motion is stable, but if the values of $m/\gamma$ is less than some critical value
the disturbed motions (\ref{sl}) are unstable. Amplitudes of small disturbances of the exact solution (\ref{sl}) grow with growing time $t$.

\begin{figure}[h]
\centering
\includegraphics[width=\textwidth]{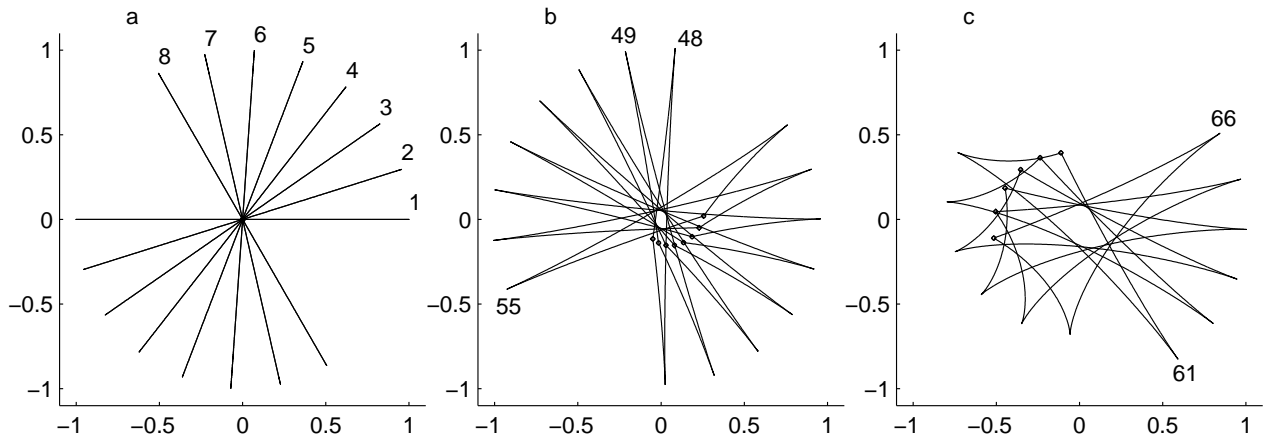}    
\caption{\small Evolution of the instability for the motion
(\ref{sl}).}
\end{figure}

The example of such a motion is shown in Fig.~2. Here
$m/\gamma=1$; $n=1$; $b_k^\mu=0$; $a_0=A_1=1$ and initial
conditions have the form \be \rho^\mu(\lm)=e_1^\mu\sin\lm,\qquad
v^\mu(\lm)=e_0^\mu+e_2^\mu(\sin\lm+0{.}001\cos^2\lm),\qquad\lm\in
[0,2\pi]. \label{init}\ee

We see that at the initial stage (Fig.~2a) the folded string uniformly rotates,
keeping the rectilinear form, and the massive point is at rest at the center.
After some time interval (depending on the amplitude of initial disturbance)
the massive point leaves the center of rotation and the string shape changes.
This stage is shown in Fig.~2b. The further evolution results in the essential change of the string shape (Fig.~2c), the string takes the form of rotating
curvilinear triangle and the massive point periodically moves from one its corner to another. This character of quasiperiodical motion is the final stage
of  any disturbed state (\ref{sl}), it does not depend on the form of initial
disturbance. But the typical time of breaking an initial state depends on
amplitude of this disturbance.
If the initial data corresponding to solutions (\ref{sl}),
are not disturbed, experiments demonstrate instability of the motion
because of disturbances from errors of numerical calculations.

However, at large values of mass $m$ or parameter $m/\gamma$ (for the given
initial conditions (\ref{sl}) this means $m/\gamma>6$)
numerical experiments shows, that the massive
point does not leave the center of rotation, and motions
(\ref{sl}) are stable.

This effect is similar to the spontaneous symmetry breaking.
The state (\ref{sl}) remains symmetric and stable if the mass of the
material point $m$  exceeds some critical value: $m>m_{cr}$.
But the symmetry of this state is breaking spontaneously
because of instability in the case
\be
m<m_{cr}.
\label{mcr}\ee

These results of numerical experiments are unexpected
and an analytical verification of them is required.
The analytical study of small disturbances of rotational states is presented
in the following section.

\bigskip

\centerline {\bf 3. Spectrum of small disturbances}
\medskip

We denote $\breve\Psi^\mu_\pm(\tau\pm\s)$ the functions in the expression
(\ref {gensol}) for the considered rotational motions (\ref {hyp}),
(\ref {rot}) and (\ref {sl}).
For example, for the world surfaces (\ref {sl}) (with the mass at the center)
derivatives of these vector functions are
\be
\breve\Psi^{\pr\mu}_\pm(\tau)=e_0^\mu
a_0\pm\sum_{k>3} e^\mu_k b_k\pm nA_n \cdot e ^\mu (n\tau)
\label{Psirot} \ee
and for the hypocycloidal states (\ref{hyp}) they are
\be \breve\Psi^{\pr\mu}_\pm (\tau)=e_0^\mu(1\mp\th)\,
a_0\pm\sum_{k>3} e^\mu_k b_k+\om_nA_n\Big[\pm(C_0-C)\,e ^\mu(\om_n\tau)+
 (S_0\mp S)\,\acute e^\mu(\om_n\tau)\Big].
\label{Psihyp} \ee
Here
\be
\th=\frac{\tau_0}{2\pi},
\label{theta}\ee
the values $\tau_0$ and $2\pi$ are roots of Eqs.~(\ref{con}), (\ref{omtaub}),
and notations (\ref{eacute}), (\ref{cscs}) are used.

To describe any small disturbances of the rotational motion of the
system, that is motions close to states (\ref{hyp}), (\ref{rot})
or (\ref{sl}) we consider vector functions $\Psi^{\pr\mu}_\pm$
close to $\breve\Psi^{\pr\mu}_\pm$ in the form
\be
\Psi^{\pr\mu}_\pm(\tau)=\breve\Psi^{\pr\mu}_\pm(\tau)
+\f_\pm^\mu(\tau).\label{Psi+f}\ee

The disturbance $\f_\pm^\mu(\tau)$ is supposed to be small, so we
omit squares of $\f_\pm$ when we substitute the expression (\ref{Psi+f})
into dynamical equations (\ref{close}) and (\ref {mg}).
In other words, we work in the first linear vicinity of the states
(\ref{hyp}), (\ref{rot}) or (\ref{sl}).
Both functions $\Psi^{\pr\mu}_\pm$ and $\breve\Psi^{\pr\mu}_\pm$ in expression
(\ref{Psi+f}) must satisfy the condition (\ref {Psi2}), hence in the first order
approximation on $\f_\pm$ the following scalar product equals zero:
\be
\big\langle\breve\Psi^\pr_\pm,\f_\pm\big\rangle=0.
 \label {Psif}
\ee

For the disturbed motions the equality (\ref {Ctau0}) $\tau^*=\tau+\tau_0$,
generally speaking, is not carried out and should be replaced with the
equality
\be
\tau^*=\tau+\tau_0+\zeta(\tau),
\label{taudel}\ee
where $\zeta (\tau) $ is a small disturbance.

Expression (\ref{Psi+f}) together with Eq.~(\ref{gensol}) is the
solution of the equations of string motion (\ref {eq}).
Therefore we can obtain the equations of evolution of small disturbances
$\f_\pm^\mu(\tau)$, substituting expressions (\ref{Psi+f}) and (\ref{taudel})
with Eqs.~(\ref{Psirot}), (\ref{Psihyp}) in two other equations of motion
(\ref{close}) and (\ref{mg}).
We take into account the nonlinear factor $\big[\dot X^2(\tau,0)\big]^{-1/2}$
and contributions from the disturbed argument $\tau^*$ (\ref{taudel}):
$$
\breve\Psi^{\pr\mu}_\pm(\tau^*\pm2\pi)\simeq\breve\Psi^{\pr\mu}_\pm(\pm)+
\breve\Psi^{\pr\pr\mu}_\pm(\pm)\,\zeta(\tau).
$$
Here and below $(\pm)\equiv(\tau+\tau_0\pm2\pi)$.

This substitution results in the following linearized system of equations
in linear (with respect to $\f_\pm^\mu$ and $\zeta$) approximation:
\be
\begin {array}{c}
\f_+^\mu(+)+\f_-^\mu(-)+2a_0\big[e_0^\mu+v\acute e^\mu(\om_n\tau)\big]\,
\zeta'(\tau)-2a_0v\om_n e^\mu(\om_n\tau)\,\zeta(\tau)=
\f_+^\mu(\tau)+\f_-^\mu(\tau),\\
\displaystyle
\frac d{d\tau}\bigg\{\f_+^\mu(\tau)+\f_-^\mu(\tau)-\frac1{1-v^2}
\big[e_0^\mu+v\acute e^\mu(\om_n\tau)\big]
\big[\f_+^0+\f_-^0+v(\acute\f_++\acute\f_-)\big]\bigg\}+
\rule{0mm}{7.5mm}\\
+Q\Big[\f_+^\mu(+)-\f_-^\mu(-)-\f_+^\mu(\tau)+\f_-^\mu(\tau)+
2\om_n^2A_n f^\mu(\tau)\cdot\zeta(\tau)\Big]=0.
\rule{0mm}{6.5mm}\end{array}
\label{sysf}\ee
Here $f^\mu(\tau)=(C-C_0)\,\acute e^\mu(\om_n\tau)+S_0 e^\mu(\om_n\tau)$
for the states (\ref{hyp}) and (\ref{rot}),
$v$ is the speed of massive point satisfying Eqs.~(\ref{a0A}), and we use
the following notations for the scalar products:
\be
\f_\pm^0\equiv\langle e_0,\f_\pm\rangle,\quad
\f_\pm^k\equiv\langle e_k,\f_\pm\rangle,\quad
\f_\pm\equiv \langle e,\f_\pm\rangle, \quad
\acute\f_\pm\equiv\langle\acute e,\f_\pm\rangle.
\label{fiscal}\ee

For the rotational state (\ref {sl}) with the  mass at the center
we are to put $\om_n=n$, the speed $v=0$
and $f^\mu(\tau)=\acute e^\mu(n\tau)$ in Eqs.~(\ref{sysf}).

If we take scalar products of equations (\ref{sysf}) onto the
basic vectors $e_0$, $e_k$, $e(\tau)$, $\acute e(\tau)$, and add
the equations (\ref{Psif}) after substituting expressions (\ref{Psihyp})
\be
(1\mp\th)\,\f_\pm^0(\tau)\pm a_0^{-1}\sum\limits_{k>3}b_k
\f_\pm^k(\tau)\mp\frac{\om_nv}{2Q}\f_\pm(\tau)+
\big(v\mp\th/v\big)\,\acute\f_\pm(\tau),
\label{sysPsif}\ee
we obtain the linear system of differential equations with respect
to projections (\ref{fiscal}) $\f_\pm^0(\tau)$, $\f_\pm^k(\tau)$,
$\f_\pm(\tau)$, $\acute\f_\pm(\tau)$, and the function $\zeta(\tau)$.
This system has constant coefficients but it also has deviating arguments
$(\pm)$ together with $(\tau)$.

We search solutions of this system in the form of harmonics
\be
\f_\pm^0=B_\pm^0 e^{-i\tom\tau},\quad \f_\pm^k=B_\pm^k e^{-i\tom\tau},
\f_\pm=B_\pm e^{-i\tom\tau},\quad \acute \f_\pm=\acute B_\pm e^{-i\tom\tau},
\quad 2a_0\zeta=\Delta e^{-i\tom\tau}.
\label{fexp} \ee

This substitution results in the linear homogeneous system of algebraic
equations with respect to the amplitudes of harmonics (\ref{fexp}).
For the rotational state (\ref {sl}) with the mass at the center
with the function $\breve\Psi^{\pr\mu}_\pm$ in the form (\ref{Psirot})
this linear system is
\be
\begin {array} {c}
B_+^kE_++B_-^kE_-= 0, \qquad B_+^k (QE_+-i\tom) =B_-^k (QE_-+i\tom); \\
B_+^0E_++B_-^0E_-= i\tom\Delta, \qquad B_+^0E_+= B_-^0E_-,\rule {0mm} {5mm} \\
B_+E_++B_-E_-= 0, \qquad\acute B_+E_++\acute B_-E_-= 0, \rule {0mm} {5mm} \\
B_+(QE_+-i\tom)-B_-(QE_-+i\tom)-n\acute B_+-n\acute B_-= 0, \rule {0mm} {5mm} \\
\acute B_+(QE_+-i\tom)-\acute B_-(QE_-+i\tom)+nB_++nB_-Qn\nu\Delta=0;
\rule {0mm} {5mm} \\
B_\pm^0\pm\nu B_\pm\pm a_0^{-1}\sum\limits_{k>3}
b_kB_\pm^k=0.\rule {0mm} {5mm}
\end {array}
\label{sysB}\ee
Here $E_\pm=\exp(\mp2\pi i\tom)-1$,
\be
\nu=\frac{nA_n}{a_0}=\Big(1-a_0^{-2}\sum\limits_{k>3}b_k^2\Big)^{1/2}.
\label{nu}\ee
The last equations in system (\ref{sysB}) are the consequence of
Eqs.~(\ref{sysPsif}).

If $b_k=0$, the first two equations of system (\ref{sysB}) form the
closed subsystem. It has non-trivial solutions if and only if
$\tom$ is a root of the equation
$$
\sin\pi\tom\cdot\Big(\sin\pi\tom+\frac{\tom}{2Q}\cos\pi\tom\Big)=0.
$$

It coincides with Eq.~(\ref{q}). Hence, the spectrum of
transversal (with respect to the $xy$ plane) small fluctuations of
the string for the state (\ref {sl}) contains the same frequencies,
as $\om_n$ from solutions (\ref {rot}) and (\ref {sl}).
All these frequencies are real numbers, therefore amplitudes of such
fluctuations do not grow with growth of time $t$.

We in the greater degree are interested in disturbances in the $xy$ plane,
shown in Fig.~2. Assuming for such fluctuations $B_\pm^k=0$,
we study the condition of existence of non-trivial solutions for the last
8 equations of the homogeneous system (\ref{sysB}).
This condition (vanishing the corresponding determinant) is reduced
to the following equation:
\be
4Q^2\tan^2\pi\tom       
+4Q\left(\tom+\frac{n^2\nu^2}{2\tom}\right)\tan\pi\tom+\tom^2-n^2=0.
\label{eqtom} \ee
This transcendental equation contains the denumerable set of real roots
(frequencies). They correspond to different modes of small oscillations
of the string in the rotational state (\ref {sl}).

This state will be unstable, if there are complex frequencies
$\tom=\om+i\xi$ in the spectrum, generated by Eq.~(\ref{eqtom}).
If its imaginary part $\xi$ will be positive, the modes of
disturbances $\f^\mu$ (corresponding to the root $\om+i\xi$) get the
multiplier $\exp(\xi\tau)$, that is they grow exponentially.

The search of complex roots of equation (\ref{eqtom}) shows that
such roots can exist only on the imaginary axis of the complex
plane $\tom$. The typical behavior of these roots is shown in
Figs.~3a and 3b. Here $b_k=0$, hence $\nu=1$. Level lines of the
real part of l.h.s. of Eq.~(\ref{eqtom}) Re$F(\om+i\xi)=0$ are
marked with black lines and zeros of the imaginary part
Im$F(\om+i\xi)=0$ --- with red lines.  Roots of Eq.~(\ref{eqtom})
correspond to cross points of these lines.

On the imaginary axis of $\tom$ (in the case $\tom=i\xi$)
the equation (\ref{eqtom}) is transformed into the form
\be                   
4Q^2\tanh^2\pi\xi +\xi^2+n^2=4Q\left(\frac{n^2\nu^2}{2\xi}-\xi\right)
\tanh\pi\xi.
\label{eqxi}\ee
The left hand side of this equation grows with growth $\xi$ (for $\xi>0$),
and the right hand side decreases. It is obvious, that the root $\xi>0$ of
Eq.~(\ref{eqxi}), that is the imaginary root $\tom=i\xi$ of Eq.~(\ref{eqtom})
exists, if and only if
\be 2\pi Q\nu^2>1.
\label {crit1} \ee

If we use the expression (\ref{a0A}) in the form
$Q=\gamma a_0/m$ (remind that $v=0$ in the case under consideration) and
Eq.~(\ref{nu}) we reduce the criterion (\ref{crit1}) to the
following form:
\be
m<m_{cr}\equiv 2\pi\gamma a_0\Big(1-a_0^{-2}\sum\limits_{k>3}b_k^2\Big).
\label{crit2}\ee

Thus, we obtain the analytical proof of the threshold effect (\ref{mcr})
observed in numerical experiments.


\begin{figure}[ht]
\unitlength=1pt \centering
\includegraphics[trim=30 350 0 160,clip]{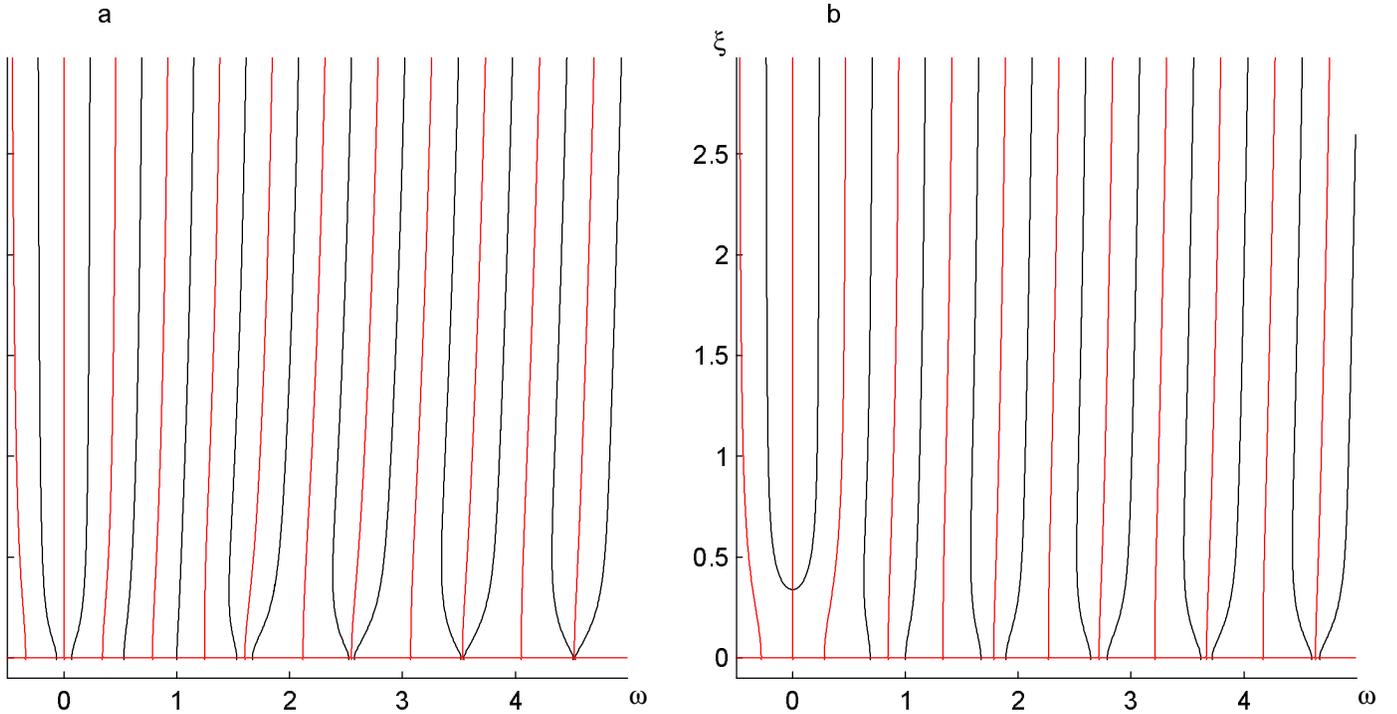}
\caption{\small Zero level lines for real and imaginary part (red)
of Eq.~(\ref{eqtom}); a) $m=6.5\gamma a_0$, b) $m=\gamma a_0$}
\end{figure}

In the particular case $b_k=0$ (that is cyclic coordinates do not ``work''
or the manifold ${\cal M}=R^{1,3}$ is the simplest one) shown in Figs.~2
and 3, the threshold value in criterion (\ref{crit2}) takes the form
$m_{cr}=2\pi\gamma a_0$. In Fig.~3a the value $m=6.5\gamma a_0$ is a bit
greater than $m_{cr}$, hence all roots
are real ones. But in Fig.~3b the inequality (\ref{crit2})
takes place, so the imaginary root $\tom=i\xi$ appears and the
corresponding amplitude of disturbances grows exponentially:
$\f=B e^{\xi\tau}$.

This threshold effect or the spontaneous symmetry breaking for the string
state (\ref{sl}) was observed in details in numerical experiments in Sect.~2.
Note that our analysis in Sect.~3 is suitable only for initial stage of
an unstable motion when disturbances are small.
The further evolution of the disturbed motion (shown in Figs.~2c)
can be described only in the numerical procedure.

\smallskip

The similar investigation of small disturbances in the form (\ref{Psi+f})
for the rotational states (\ref{hyp}) and (\ref{rot}) includes
substitution of harmonics (\ref{fexp}) into the linearized system
of differential equations for projections (\ref{fiscal}) of $\f_\pm^mu$.
In this case the first two equations of system (\ref{sysB}) keep
the same form
$$
B_+^kE_++B_-^kE_-= 0, \qquad B_+^k (QE_+-i\tom) =B_-^k (QE_-+i\tom);
$$
but here $E_\pm=\exp\big[-i(\tau_0\pm2\pi)\tom\big]-1$.
This subsystem  has non-trivial solutions if and only if
$\tom$ is a root of the equation
$$
\cos2\pi\tom-\cos\tau_0\tom=\frac{\tom}{2Q}
\sin2\pi\tom,
$$
coinciding with Eq.~(\ref{con}). Hence, in this case of small transversal
fluctuations the spectrum of roots $\tom$ also looks like the spectrum
of frequencies $\om_n$.

For the last 8 equations of the homogeneous system, generalizing (\ref{sysB}),
the condition of existence of non-trivial solutions (oscillations in $xy$
plane) for the states (\ref{rot}) may be reduced to the equation,
decomposing into product of the following two equations:
\be
2Q\tom\Big[v^2-C+(1-v^2)\cos2\pi\tom\Big]+\om_n^2v^2\sin2\pi\tom=0,
\label{osc1}\ee
\vspace{-3mm}
\be
\Big[\om_n^2+4Q^2(1-v^2)\Big](1-\cos2\pi\tom)+
4Q(1-v^2)\,\tom\sin2\pi\tom+\tom^2\Big[C-v^2+(1-v^2)\cos2\pi\tom\Big]=0,
\label{osc2}\ee

The analysis of their roots shows that all roots of Eq.~(\ref{osc1}) and
Eq.~(\ref{osc2}) are real numbers, if all values satisfy natural physical
restrictions, for example, $v<1$, $m>0$. The typical picture of these roots
is presented in Fig.~4a for Eq.~(\ref{osc1}) and in Fig.~4b for
Eq.~(\ref{osc2}).

\begin{figure}[h]
\unitlength=1pt \centering
\includegraphics[trim=30 370 0 160,clip]{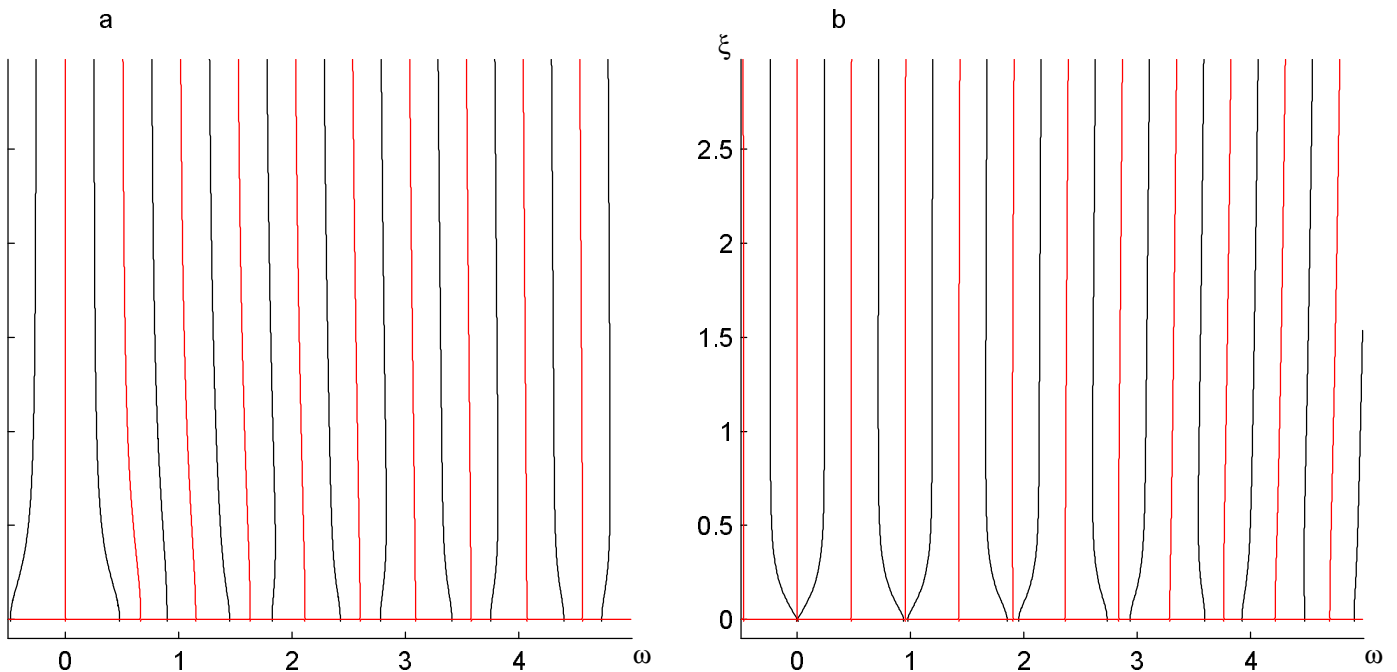}
\caption{\small Zero level lines for real (black) and imaginary
part (red) for a) Eq.~(\ref{osc1}); b) Eq.~(\ref{osc2})}
\end{figure}

Here the values of the parameters for the rotational state (\ref{rot})
are: $\om_n=0.9$, $Q\simeq1.385$, $v^2\simeq0.875$,
$(\gamma/m)^2\sum b_k^2=0.5$. We may conclude that in linear approximation
this rotational state is stable with respect to small disturbances.

\bigskip

\centerline {\bf Conclusion}
\medskip

The analysis of the stability problem for rotational states
(\ref{hyp}), (\ref{rot}) and (\ref{sl}) of the closed string with a
point-like mass was made with two different approaches in Sects.~2 and 3.
This analysis showed that the states of the class (\ref{sl}) (with the
massive point at the rotational center) are unstable if the mass
is less than the threshold $m_{cr}$ (\ref{crit2}). This effect of
the spontaneous symmetry breaking for these string states
restrains applicability of these states in hadron spectroscopy.

But our investigation of rotational states (\ref{hyp}), (\ref{rot})
showed, that they are stable in the mentioned sense.
These rotational states generate non-trivial spectrum of Regge trajectories
\cite{MilSh}.
So we can apply them for describing excited hadrons with exotic properties,
in particular, glueballs, hybrids, pentaquarks
in accordance with applications of various string hadron models
\cite{4B}, \cite{InSh} in meson and baryon spectroscopy.

\medskip

Acknowledgments

Authors are grateful for support by the RFBR grant 05-02-16722.

\end{document}